\documentclass[twocolumn,showpacs,superscriptaddress,preprintnumbers,amsmath,amssymb,prl]{revtex4}
\usepackage{graphicx}
\usepackage{dcolumn}
\usepackage{bm}
\begin{document}

\title{Jamming, Yielding and Rheology of Weakly Vibrated Granular Media}

\author{Joshua A. Dijksman}
\affiliation{Kamerlingh Onnes Lab, Universiteit Leiden, Postbus
9504, 2300 RA Leiden, The Netherlands}\affiliation{Dept. of
Physics, Duke University, Science Drive, Durham NC 27708-0305,
USA}

\author{Geert H. Wortel}
\affiliation{Kamerlingh Onnes Lab, Universiteit Leiden, Postbus
9504, 2300 RA Leiden, The Netherlands}

\author{Louwrens T. H. van Dellen}
\affiliation{Kamerlingh Onnes Lab, Universiteit Leiden, Postbus
9504, 2300 RA Leiden, The Netherlands}

\author{Olivier Dauchot}
\affiliation{CEA-Saclay, SPEC-GIT, URA 2464, 91191 Gif-sur-Yvette,
France}

\author{Martin van Hecke}
\affiliation{Kamerlingh Onnes Lab, Universiteit Leiden, Postbus
9504, 2300 RA Leiden, The Netherlands}

\date{\today}

\begin{abstract}
We establish that the rheological curve of dry granular media is
non-monotonic, both in the presence and absence of external
mechanical agitations. In the presence of weak vibrations, the
non-monotonic flow curves govern a hysteretic transition between
slow but steady and fast, inertial flows. In the absence of
vibrations, the non-monotonic flow curve governs the yielding
behavior of granular media. Finally, we show that non-monotonic
flow curves can be seen in at least two different flow geometries
and for several granular materials.
\end{abstract}

\maketitle

Granular media are collections of macroscopic and athermal grains
which interact through dissipative, frictional contact forces and
that jam in metastable configurations
\cite{1995_book_duran,1996_revmodphys_jaeger,2004_epje_gdrmidi,2008_annurevfluid_forterre}.
Under sufficient shear stress, or when mechanically agitated,
granular media yield and flow~\cite{2010_prl_nichol,
2007_pre_sanchez, 2006_pre_rubin,1989_prl_jaeger,
2009_jrheol_marchal, 2009_epl_janda, newpouliquen}. Here we ask:
what is the rheological scenario that connects these observations?

The classical example of a inclined layer of sand illustrates that
granular media exhibit a finite flow threshold, and  once the
material yields, the flow rate jumps to a finite value
\cite{2008_annurevfluid_forterre,1980_powtech_nedderman,
1995_book_duran, 2004_epje_gdrmidi}. The simplest flow scenario
that captures this behavior is sketched in Fig.~1a, where in
analogy to static and dynamic friction, the static yield stress
$\sigma_s$ is assumed to be larger  than the dynamic yield stress
$\sigma_d$.

Such scenario implies that the flow rate continuously decreases to
zero when the stress (inclination angle) is lowered, as can be
seen from following the flow curve in Fig.~1a. In experiments,
however, the flow is found to stop discontinuously:
stress-controlled granular flows have a minimal flow rate
\cite{2005_book_coussot, 2002_pre_dacruz, 2008_epl_mills}. This
calls for a more elaborate flow curve than the one in Fig. 1a. We
sketch one that was proposed earlier~\cite{2005_book_coussot,
jaeger1990} in Fig.~1b, where the negative slope, signalling an
instability, leads to a ``forbidden'' range of flow rates --- when
the stress is lowered below $\sigma_{min}$, the flow rate jumps to
zero~\cite{2005_book_coussot, jaeger1990}.

In this Letter, we firmly establish the existence of non-monotonic
flow curves for granular media by probing their rheology, both in
the presence and absence of externally supplied vibrations of
strength $\Gamma$.
%
In the absence of vibrations $(\Gamma\!=\!0)$, the flow curves
indeed are of the form as sketched in Fig.~1b, i.e., with a finite
yield stress, and a dip at intermediate flow rates. As we will
see, the rheology at $\Gamma\!=\!0$ can be seen as a limiting case
of the more general rheology at finite vibration strength. For
$\Gamma\!>\!0$ we find flow curves as sketched in Fig.~1c, i.e.,
without a clear yield stress and with two competing branches of
positive slope. This lower branch is consistent with recent
experiments which have found that external agitations decrease or
even quash the yield stress of granular materials
\cite{2010_prl_nichol, 2007_pre_sanchez,
2006_pre_rubin,1989_prl_jaeger, 2009_jrheol_marchal,
2009_epl_janda, newpouliquen}, allowing for very slow flows under
constant, small, stresses and finite vibration strength.
Non-monotonic flow curves such as depicted in Fig.~1c are
well-known to arise for polymer melts, micelles and viscous
suspensions \cite{2003_annurevfluid_goddard,schall,
2005_book_coussot, 2007_pre_heymann} but have not been observed,
to the best of our knowledge, for dry granular media.

\begin{figure}[t!]
\begin{center}
\includegraphics[width=8cm]{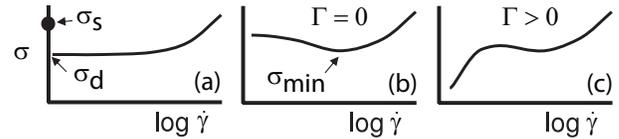}
\caption{Hypothetical flow curves for granular media, relating
strain rate $\dot{\gamma}$ and stress $\sigma$. (a) Monotonic flow
curve with a different static ($\sigma_s$) and dynamic
($\sigma_d$) yield stress. (b) Non-monotonic flow curve - the
region with negative slope signals 'forbidden' strain rates in
stress controlled experiments. (c)  A typical flow curve observed
in our experiments for finite vibration strengths. } \label{fig1}
\end{center}
\end{figure}

The non-monotonic flow-curves are obtained in experiments with
controlled flow-rates, and negative slopes are expected to cause
instabilities and hysteresis when the stress is controlled
\cite{schall}. For $0\!<\!\Gamma\!\lesssim\!1$, we find that
stress sweeps through the unstable regime lead to a concomitant
hysteretic transition of the flow rate between the two stable
branches. For $\Gamma\!=\!0$, we find that stress-controlled
yielding, i.e., the hysteretic transition from zero flow to finite
flow rate, is intimately connected to the dip in the
$\Gamma\!=\!0$ flow-curve. We finally show that non-monotonic flow
curves like in Fig.~1b can be seen independent of flow geometry
and for several, but not all, granular materials. Our findings of
the robust connection between non-monotonic flow curves and
rheological instabilities shine new light on the nature of the
granular jamming and yielding transitions.

\begin{figure}[t!]
\begin{center}
\includegraphics[width=8cm,viewport=00 00 350 230,clip]{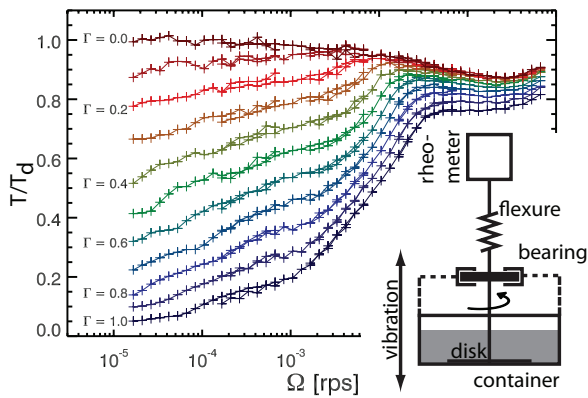}
\caption{(Color online) Flow curves $T(\Omega)$ for
$\Gamma=0,0.1,\dots 1$ as indicated. We normalize $T$ by the
dynamic yield torque $T_d$. Inset: vibrated split-bottom
rheological setup.
 } \label{fig1'}
\end{center}
\end{figure}

{\em Setup  ---} Our experiments are carried out in a split-bottom
shear cell (Fig.~\ref{fig1'}),  in which a layer of glass beads
(diameter 1-1.3 mm) of depth $H = 18$ mm is driven by the rotation
of a rough disk of radius $R_s=4$ cm mounted flush with the
bottom. This flow geometry has been studied extensively and
produces smooth, robust and well-controlled granular flows
\cite{fenistein,2006_prl_cheng, 2010_pre_dijksman}.

An important novel aspect of our setup is that the shear cell can
be vibrated vertically and sinusoidally (distortion $< 1$\%) with
frequency $f$ and amplitude $A$. We fix $f=63$ Hz in the middle of
a frequency window where no mechanical resonances arise. Strictly
vertical vibration is ensured by guiding the motion of the shear
cell with a levelled square air-bearing (4"x4", New Way) which is
coupled to a electromagnetic shaker (VTS systems VG100). We
control the dimensionless shaking strength, defined as $\Gamma:= A
(2\pi f)^2/g$, with a feedback loop to within $<10^{-3}$.

We also control the rotation rate, $\Omega$, and applied torque,
$T$, by a rheometer (Anton Paar DSR 301), which is coupled to the
vibrating cell by means of a flexure with a torsional spring
constant of 4 Nm/rad and compressional spring constant of $5
\!\times \!10^2$ N/m. We perform rheological experiments at fixed
$\Gamma$ and either control the torque $T$ and measure the
resulting rotation rate $\Omega$ or vice-versa. All flow
experiments are preceded by appropriate pre-shear. Disk rotation
is always continuous; stick-slip is not observed. Note that, as in
other flow geometries, the local strain rate and stress in the
split bottom cell vary throughout the cell \cite{2007_epl_depken}.
We thus probe the grain rheology with $T$ as a proxy for the
stress $\sigma$, and $\Omega$ as a proxy for the strain rate
\cite{2010_pre_dijksman, 2002_pre_dacruz}. Hence, the
experimentally observed curves for $T(\Omega)$ are best thought of
as global flow curves.

{\em Main Phenomenology: Flow curves ---} Fig.~2a shows the flow
curves $T(\Omega)$, determined in experiments in which the
rotation rate $\Omega$ is controlled, and the average torque $T$
is measured (after removing transients).

The flow curve for $\Gamma\!=\!0$ is non-monotonic. For small flow
rates ($\Omega < 10^{-3}$ rad/s), the stress reaches a plateau
from which we determine the dynamical yield torque $T_d$ as $13.9
\pm 0.1$ mNm --- this value is set by the
geometry~\cite{fenistein} and the effective friction coefficient
of the grains. For increasing $\Omega$, $T$ decreases until it
reaches a minimum torque $T_{min}$ of about $12.1 \pm 0.1$ mNm at
$\Omega \approx 0.3$ rps. This non-monotonic effect is substantial
in magnitude and has not been observed for granular flows before.
Around the minimum, the inertial number near the split is of order
one, and we associate the increase of torque for larger rates with
the onset of inertial
flows~\cite{2004_epje_gdrmidi,2008_annurevfluid_forterre}.

The flow curves for $\Gamma > 0$ exhibit similarly non-monotonic
behavior, but differ for small $\Omega$. As shown in Fig.~2a, the
flow rates over which the flow curves have negative slope become
smaller for larger $\Gamma$. At the lower $\Omega$ range of this
regime, $T(\Omega)$ reaches a local maximum and for even smaller
$\Omega \lesssim 0.02$ rps, we observe a decrease of $T$ with
$\Omega$ as $T \sim log(\Omega)$. In additional experiments at
fixed $T$ we have carefully checked that the flow is stable and
steady in this positive slope regime. This regime only exists for
finite agitation strength and signals a novel flow regime of
mechanically agitated granular flows which is unique to
$\Gamma>0$, as suggested in Fig.~1c.

We conclude that our flow curves are consistent with the flow
scenarios depicted in Fig.~1b-c. In other systems with
non-monotonic flow curves, fixing the flow rate in the
negative-slope regime typically leads to a separation of the
system into two regimes, one with low, and one with large
strain-rate: in other words, shear banding \cite{schall}. In
contrast, we have not seen any clear evidence for such behavior in
our system --- the flow profiles as observed at the free surface
do not appear to change when we fix the flow rate in the
negative-slope regime. We note that the standard shear banding
mechanism depends on the shear stresses being sufficiently
homogeneous, while in our system we have a strongly inhomogeneous
stress field emanating from the split in the bottom
\cite{2007_epl_depken}
--- this inhomogeneity is crucial in obtaining a smooth granular
flow, but may hinder the observation of additional shear banding.
Our set up does not allow us to determine whether dilatancy plays
a significant role in the development of the non-monotonic flow
curves. Yet, since we only consider slow steady flows, the usual
inertial and transient dilatancy
effects~\cite{2008_annurevfluid_forterre, pailha2008} are
certainly ruled out.

\begin{figure}[t!]
\begin{center}
\includegraphics[height=8cm,angle =90, viewport=150 50 510 815,clip]{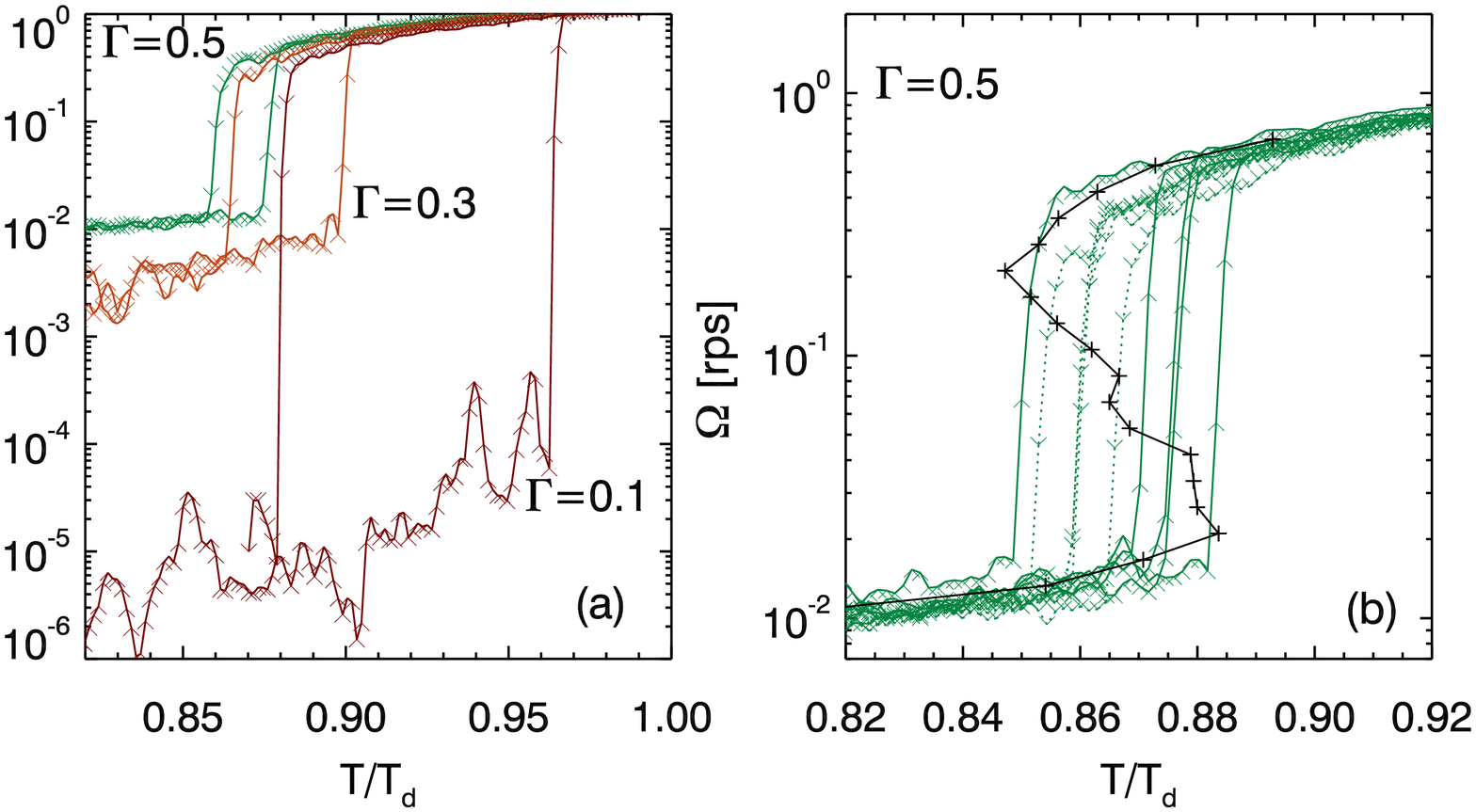}
\caption{(Color online) (a) Finite $\Gamma$ hysteresis loops for
$\Gamma = 0.1, 0.3, 0.5$. (b) Several hysteresis loops at $\Gamma
= 0.5$, with the rheological data from $\Gamma= 0.5$ from Fig.~2
overplotted in black.} \label{fig:4}
\end{center}
\end{figure}

{\em Rheological instability for $\Gamma>0$ ---} We now turn our
attention to torque-controlled experiments, and will probe whether
the negative sloped regime of the flow curves for $\Gamma>0$ leads
to hysteresis. To do so, we slowly ramp $T/T_d$ up and down
between 0.8 and 1.1, i.e., through the multi-valued regime.
Fig.~\ref{fig:4}a illustrates the resulting hysteresis loops.
Ramping upwards, we observe a sudden jump from the slow,
mechanically agitated flow branch to the rapid, inertial branch.
Ramping downwards makes the flow rate jump back to the slow,
mechanically agitated branch --- there is considerable hysteresis
between the stresses where these jumps happens. For smaller
$\Gamma$, the gap between slow and rapid flow rates increases,
consistent with the flow curves shown in Fig.~2.

In Fig.~\ref{fig:4}b we further strengthen the direct connection
between the negative slope of the $T(\Omega)$ curve and the
hysteresis observed in the $\Omega(T)$ curves for the example of
$\Gamma\!=\!0.5$. We combine several torque controlled data sets
with the appropriate flow curve, and observe that while the
precise location of individual hysteresis loops fluctuates, the
characteristic torques remain confined to an interval which
coincides well with the minimum and maximum of the $T(\Omega)$
curves. We concluded that for $\Gamma > 0$, hysteresis and
negatively sloped flow curves are directly related.

{\em Static yield vs dynamic yield for $\Gamma = 0$ ---} In the
absence of vibrations, the connection between flow curves and
instabilities is more subtle, as there is only one stable branch
with a finite flow rate (although one could see the jammed state
at small stress as the second ``stable branch''). We now ask the
following question: when we slowly ramp up the torque, how is the
resulting  yielding behavior influenced by the non-monotonic flow
curve?

\begin{figure}[t!]
    \begin{center}
        \includegraphics[angle= 90, width=8cm, viewport=0 340 520 800, clip = true]{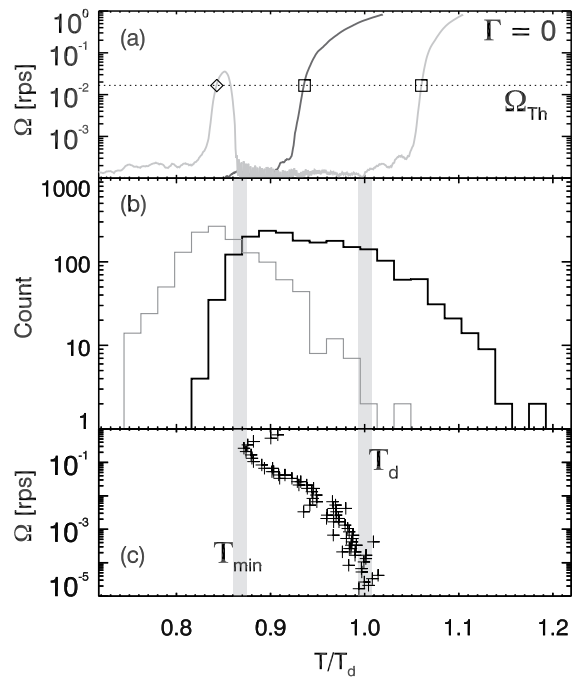}
        \caption{(a) Typical examples of $\Omega(T)$ curves during
        a torque ramp. The dashed line indicates the threshold rotation rate
        $\Omega_{Th} = 0.016$ rps that defines a yield event. $\diamond$ indicates a
        micro yielding event, $\Box$ a global failure event. (b) The
        probability distribution function of micro yielding (grey) and
        global yielding (black). Grey bars indicate the onset value for the
        rheological instability $T_{min}$, and the dynamic yield stress
        $T_d$. (c) Flow curve for $\Gamma=0$.  \label{fig2}}
\end{center}
\end{figure}

To answer, we study the statistics of static yielding by ramping
up the applied torque at 0.5 mNm/s, fixing $\Gamma = 0$, and
measuring the ensuing $\Omega(t)$. We identify a yield event
whenever $\Omega > \Omega_{Th} = 0.016$ rps. Our statistics are
robust for $\Omega_{Th}$ between 0.002 and 0.13 rps, and over the
duration of the experiment ($\sim$ 10h), we do not observe any
appreciable drift in the properties of the distribution. As
indicated in Fig.~\ref{fig2}a, we observe two types of yielding:
micro yielding, where $\Omega$ only briefly peaks above
$\Omega_{Th}$ (diamond), and global failure, where the increase in
$\Omega$ is dramatic and persistent (square).

We measure the statistical properties of both types of yielding
over 1943 torque ramps. In Fig.~\ref{fig2}b we show the
probability distribution functions of the micro yielding torques
(gray) and the global yielding torques (black). Consistent with
the flow curve for $\Gamma = 0$ (Fig.~3c), we observe that micro
yield events do not occur above $T/T_d\!=\!1$ --- once the torque
is above the dynamic yield threshold, the material flows at a rate
given by the positive slope region of the flow curve. What is
surprising is that global yielding can happen both {\em below} and
{\em above} $T_d$
--- invalidating the simple picture of a lower dynamic, and higher
static yield threshold. Rather, the probability for complete
failure is bounded by $T = T_{min}$ --- for lower torques the flow
curve shows there is no steady state flow, and  only micro
yielding events are observed.

We conclude that, due to the non-monotonic flow curve, there are
three yielding regimes. Below $T_{min}$, micro yielding without
global yielding; above $T_{min}$ but below $T_d$, micro yielding
with a finite probability of global yielding; above $T = T_d$ all
yield events become global. This intermediate regime is clearly
inconsistent with the simple picture where hysteresis in jamming
and yielding of granular media is explained from the static
threshold exceeding the dynamic one (Fig.~1a).

\begin{figure}[t!]
\includegraphics[viewport=00 000 330 160,clip,width=7.5cm]{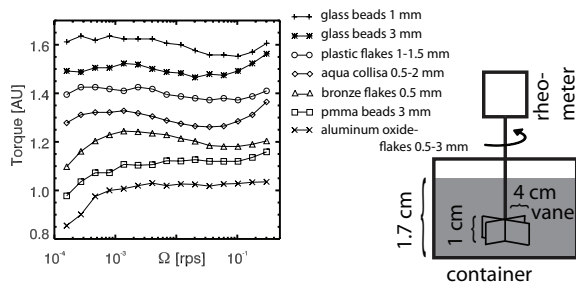}
\caption{Rheological curves for a range of materials as indicated,
where a vane (see sketch) directly coupled to an Anton Paar MCR501
rheometer was placed just above the bottom. All torques are
normalized by their mean value and (with the exception of the
bottom curve) shifted upwards for clarity.}
 \label{fig3}
\end{figure}

{\em Robustness non-monotonic flow curves ---} How specific to the
precise flow geometry and materials used are the non-monotonic
flow curves we observe? We test this in two ways. First, we can
test the dependence of the rheology on the shape of the shear
bands in the split bottom cell itself with the glass beads
mentioned earlier: for larger filling height, i.e. $H/R_s = 0.7$,
the flow field is qualitatively very different
\cite{fenistein,2006_prl_cheng, 2010_pre_dijksman}. We have
conducted a series of experiments and we observe qualitatively the
same flow curves and instabilities (not shown here).

Second, we probed the existence of the non-monotonic flow curves
for a range of materials in a standard vane geometry at $\Gamma =
0$. The top curve shown in Fig.~\ref{fig3} shows that the
non-monotonic flow curves for glass beads at $\Gamma = 0$ are not
specific to the split-bottom geometry. However, details of the
flow curves are material dependent - some materials (aluminum
flakes and PMMA beads) do not have a negative slope at all, while
others (bronze flakes, aqua collisa) have the
positive/negative/positive slope combination that we saw above for
$\Gamma > 0$ only. We do not know the cause of this material
dependence, but suggest that material dependent plastic flow in
the contact asperities may destroy the negative slope, and that
for such materials, temperature may play a similar role as
vibrations does for glass beads.

{\em Conclusion ---} We have shown that non-monotonic flow curves
are a  robust feature of slow granular flows. As a consequence,
the yielding transition at $\Gamma=0$ exhibits all the hallmarks
of a 1st order or subcritical transition. For finite $\Gamma$, the
yield stress vanishes but the hysteretic nature of the transition
persists up to a large value of $\Gamma$ (around 1 for this
filling height, somewhat smaller for larger $H/R_s$), where the
transition becomes 2nd order and smooth.

Our data suggest that material properties are key, and while we
cannot exclude collective effects \cite{Hartley,Behringer}, a
simple decrease of the particle-particle dynamic friction
coefficient with rate may be sufficient \cite{Baumberger}. Our
findings provide crucial input for refining and extending current
descriptions of granular flows
\cite{2004_epje_gdrmidi,2008_annurevfluid_forterre}.

Acknowledgement: JAD acknowledges funding from the Dutch physics
foundation FOM, and OD from the KNAW. We thank Bruno Andreotti,
Bob Behringer and Eric Cl\'ement for discussions and Jeroen Mesman
for outstanding technical assistance in the construction of the
rheological setup.



\end{document}